# Time reversal symmetry breaking and zero magnetic field Josephson diode effect in Dirac semimetal $Cd_3As_2$-mediated asymmetric SQUIDs

W. Yu[1], J.J. Cuozzo[2], K. Sapkota[1], E. Rossi[3], D.X. Rademacher[1], T.M. Nenoff[1], W. Pan[2]

[1] Sandia National Labs, Albuquerque, NM 87185, USA

[2] Sandia National Labs, Livermore, CA 94551, USA

[3] Department of Physics, William & Mary, Williamsburg, VA 23187, USA

**Abstract**

A *zero-magnetic-field* Josephson diode effect (JDE) is observed in an asymmetric superconducting quantum interference device (SQUID) mediated by Dirac semimetal $Cd_3As_2$. We argue that a phase coupling between the surface and bulk superconducting channels, a unique phenomenon recently identified in the observations of fractional Josephson effect and Leggett modes in $Cd_3As_2$, can break time reversal symmetry and therefore give rise to the zero-field JDE. Our results are anticipated to have important implications in superconducting electronic circuit applications.

The diode effect in a p-n junction plays an important role in modern microelectronics. Due to broken inversion symmetry between the electron (n) and hole (p) doped regimes, electronic transport is non-reciprocal, i.e., electrical current can flow only in one direction. This non-reciprocal nature has been widely utilized in electronic devices such as transistors, light-emitting diodes, solar cells, etc.

Recently, a similar diode effect has attracted a great deal of interest in superconducting systems [1-66]. Like the diode effect in the p-n junctions, the superconducting diode effect (SDE), or specifically the Josephson diode effect (JDE) in Josephson junctions (JJs), is expected to find important applications such as passive on-chip gyrators and circulators [66]. Such devices would be particularly impactful in quantum computing applications. Moreover, the SDE/JDE can be utilized as an alternative method to study novel superconductor properties, such as finite momentum Cooper pairing [2,10].

In a typical JJ or superconducting quantum interference device (SQUID), the I-V curve is linear in the high current regime where the device is in the normal state, see Figure 1d. The voltage $V_{dc}$ drops abruptly to zero at the so-called retrapping current $I_{+r}$ (for the current sweeping down) and stays at zero over a large current range until a switching current $-I_{-c}$ is reached. Herein, we take this switching current as the critical current ($I_c$) of the JJ and use the terminology of critical current throughout in the paper. Beyond $-I_{-c}$, the I-V curve becomes linear, and the device enters the normal state again. For the current sweeping up curve, a similar shape in the I-V curve is observed, and the positions of corresponding $-I_{-r}$ and $I_{+c}$ are marked. In general, $I_{+c} = I_{-c}$, independent of current sweeping directions as long as either time reversal symmetry (TRS) or inversion symmetry is present. However, when both symmetries are broken, the critical current can display different values depending on which direction the current is swept, a phenomenon called the JDE [1,2]. Inversion symmetry is broken in non-centrosymmetric superconducting systems or in device structures such as asymmetric SQUIDs as we will discuss in this work. However, superconductors with intrinsic broken TRS are rare because superconducting condensates are typically formed by pairs of electrons related by TRS. Consequently, external magnetic (B) fields [3] or magnetic heterostructures [4,5] are exploited to break TRS. In this regard, it is surprising that in recent experiments the JDE was observed at zero B field in non-magnetic materials [6,7,62], thus calling for more investigations into Josephson junctions made of topological quantum materials [67] where non-trivial band topology and topological superconductivity are shown to facilitate the JDE [9-19].

In this paper, we demonstrate JDE in two asymmetric SQUIDs made of Dirac semimetal [68,69] $Cd_3As_2$, see Fig. 1. In SQUID1, the efficiency $\eta_c$ of JDE, defined by $\eta_c = (I_{+c} - I_{-c})/(I_{+c} + I_{-c})$, is very weak and essentially zero at zero B field. It becomes finite at finite B fields. In SQUID2, surprisingly, a large JDE is observed at zero B field, with an efficiency of ~ 9%. Our analysis based on the resistively shunted junction model suggests that a definitive phase coupling between the surface and bulk superconducting channels can break TRS at zero B field. This, together with the geometric difference in the two JJ arms in asymmetric SQUIDs, is responsible for the observation of the zero B field JDE. Furthermore, our theoretical simulation shows that $\eta_c$ depends on the geometry of SQUIDs, thus providing an explanation of non-observation of zero-field JDE in SQUID1.

Figs. 1a and 1b shows the scanning electron microscope images of the two SQUIDs we studied in this work (see the Supplemental Material (SM) for details [70]). For electronic measurements, the devices are cool-down and immersed in a cryogenic liquid at a temperature of ~ 0.2 or 0.03 K, well below the proximity-effect induced and Al superconducting transition temperatures.

**SQUID1 Results.** I-V curves measured in SQUID1 at T = 0.2 K at B = 0 and 16 mT are shown in Fig. 1c and Fig. 1d, respectively. In both cases, current is swept first from – 7 μA to + 7 μA and then from + 7 μA to – 7 μA. Multiple switching behavior is observed for both $I_c$ and $I_r$, see inset of Fig. 1c. Here, we use the first switching position to define $I_c$ and $I_r$. This multiple switching behavior is suppressed under a finite B field, for example, at 16 mT (Fig. 1d). Moreover, at B = 0, the current up and down traces overlap almost perfectly (baring the multiple switching events), and $I_c$ and $I_r$ are close to each other for either current direction. However, at a finite B field, e.g. B = 16 mT, $I_c$ and $I_r$ differ considerably in the same current sweeping trace. For example, $I_{+r}$ = 3.8 μA, while $I_{-c}$ = 5.82 μA in the current sweeping down trace.

Differential resistance dV/dI [70] is measured together with the I-V curves. The two-dimensional (2D) color plots of dV/dI in SQUID1 as a function of DC current and B field for current sweeping up and down are shown in Figures 2a and 2b, respectively. The uniformly colored blue area in the middle represents the proximity [71] induced supercurrent regime [72-76], and the sharp edge of the region highlights the value of $I_c$ (as well as $I_r$). The B field dependence of $I_c$ (as well as $I_r$) contrasts with the typical oscillatory pattern expected in a conventional SQUID where $I_c$ oscillates with the B fields with a period inversely proportional to the area of the SQUID ring. Instead, $I_{\pm c}$ (as well as $I_{\pm r}$) displays a non-oscillatory, non-monotonic magnetic field dependence. Starting from B = 0, $I_{\pm c}$ and $I_{\pm r}$ first increase with increasing magnetic field strength, reach a maximal value which depends on the current sweep direction, and then decrease with further increasing magnetic fields. Eventually, both $I_{\pm c}$ and $I_{\pm r}$ become zero when the B field reaches the critical magnetic field (~ +/- 35 mT [72]) of the superconducting Al thin film, at which the proximity effect disappears. The enhancement of $I_{\pm c}$ with B around the zero field is strikingly large, it reaches 214% of its zero-field value at B ~ 11-12 mT. We notice here that, interestingly, this maximum enhancement is similar to what is reported in Ref. [77]. The large increase in $I_c$ is dramatically different from that in a single JJ also made of $Cd_3As_2$ where the maximal enhancement was merely 4% [72]. On the other hand, we notice that similar magnetic field response of $I_c$ (and $I_r$) was also observed [19,78-80] in the past in JJs with broken TRS and was attributed to the existence of two supercurrents that are out of phase with each other [19]. The same mechanism should also be responsible for the observed magnetic field response in our device.

Magnetic field responses of $I_{+c}$ and $-I_{-c}$ are plotted for SQUID1 in Figure 2c. The efficiency of $I_c$ JDE $\eta_c$ is plotted in Figures 2d. At B = 0, $\eta_c$ is very weak and essentially zero. As B increases, $\eta_c$ becomes finite and displays an even function of B field dependence around B = 0. At high B fields in the negative and positive directions, different B-dependent behaviors are observed. In the negative B field regime, $\eta_c$ saturates to a value of ~ 2.5% beyond -5 mT. On the other hand, in the positive B field regime, after reaching a local maximum of ~ 2.5% at B ~ 5mT, $\eta_c$ starts decreasing, and becomes negative for B > 10 mT. The magnitude of $\eta_c$ continues to increase beyond ~ 20 mT when the differential resistance become non-zero over the whole DC current range, as indicated by the color change in Figs. 2a and 2b. Overall, $\eta_c$ displays a diode-like behavior with the B field. This might be due to the interplay of an external magnetic field and the possible spontaneous breaking of time-reversal symmetry in Al-$Cd_3As_2$ heterostructures, but its exact origin is presently not known.

The B field dependence of retrapping current ($I_{+r}$ and -$I_{-r}$) and its efficiency $\eta_r = (I_{+r}-I_{-r})/(I_{+r} + I_{-r})$ are shown in Figures 2e and 2f, respectively. $\eta_r$ also shows a diode-like behavior with the B field. $\eta_r \sim 0$ between B = -20 and B = 10 mT, and then becomes finite and increases with increasing B fields. $I_r$ depends on the dissipative current in the normal state. Non-reciprocal $I_r$ represents another aspect of the JDE and has also been observed in recent experiments [5, 10, 49]. It has been shown that in the dissipative regime in the presence of a strong spin-orbit coupling, which is known to exist in $Cd_3As_2$, $I_{+r} \neq I_{-r}$ can occur [1]. It is interesting that both $\eta_c$ and $\eta_r$ follow a similar diode-like behavior with the B fields. Future studies will be focused on understanding this apparent correlation.

**SQUID2 Results.** In this SQUID, a large *magnetic-field-free* JDE is observed. The I-V curves at B = 0 T for DC currents sweeping up and sweeping down are shown in Figure 3a. For the trace of current swept down from 6 to -6 μA, $I_{-c}$ = 2.99 μA. For the trace of current sweeping up, $I_{+c}$ = 3.62 μA. This value is significantly larger than $I_{-c}$ ($I_{+c} - I_{-c}$ = 0.63 μA) and the corresponding JDE efficiency $\eta_c = (I_{+c}-I_{-c})/(I_{+c}+I_{-c})$ = 9.5%, demonstrating a large magnetic-field-free JDE in SQUID2. Like in SQUID1, there are multiple retrapping events, such as the one at -3.02 μA and the other at -2.93 μA in the sweeping up trace. A large JDE is again observed in the I-V data taken at B = 6 mT, see Figure 3b: $I_{+c}$ = 4.19 μA and $I_{-c}$= 3.50 μA. Consequently, $I_{+c} - I_{-c}$ = 0.69 μA and $\eta_c$ = 9.0%. These values are close to those at B = 0. No diode effect is observed in $I_r$, though. At B = 0, $I_{+r}$ and $I_{-r}$ appear to be equal, ~ 3 μA.

Differential resistance in SQUID2 at T = 30 mK as a function of $I_{dc}$ and B field is plotted for both current sweeping up and down traces in Fig. 4a and Fig. 4b, respectively. We note here that for the sweeping down traces, the current stops at $I_{dc} = 0$. Consequently, information on $I_{-c}$ is not available. There are several features worth pointing out. First, the non-monotonic B field dependence of the critical current is also seen in SQUID2, as in SQUID1. At B = 0, $I_{+c} = 4$ μA and it increases with increasing B fields and reaches a maximal value of ~5 μA at B = 9 mT. $I_{+c}$ then decreases with B further increased. Second, $I_{\pm r}$ also shows a magnetic field-induced enhancement. Third, there are re-entrant supercurrent states at high B fields (see more data in the SM [70]).

In Fig. 4c, $I_{-r}$, $I_{+r}$ and $\Delta I_r = I_{+r} - I_{-r}$ are plotted as a function of magnetic field. Unlike in SQUID1 where $\Delta I_r$ shows a monotonic B field dependence, $\Delta I_r$ in SQUID2 shows a much richer B field dependent behavior. At zero and small positive B fields, $I_{+r}$ and $I_{-r}$ are equal, $\Delta I_r = 0$. This reciprocal behavior persists up to 10mT. Between 10 and 17 mT, $I_{+r}$ and $I_{-r}$ generally differ from each other and $\Delta I_r \neq 0$, except at B = 12 mT, when $\Delta I_r$ quickly drops to zero. $\Delta I_r$ assumes a negative value between 17 and ~26 mT, in which there are no supercurrent states, before returning to zero again at B > 26 mT when the device enters the reentrant supercurrents regime.

The observation of a zero-B field JDE in SQUID2 is striking, as $Cd_3As_2$ itself is non-magnetic. In the following, we first argue that mechanisms, such as the Meissner [61] and circuit inductance effects [81], cannot be responsible for the JDE observed in our $Cd_3As_2$ SQUIDs. Next, we propose a time-reversal-symmetry breaking mechanism to be responsible for the existence of the zero-magnetic field JDE.

Recently, the Meissner effect was shown to play an important role in the magnetic field induced diode effect [61]. However, this effect cannot account for the JDE in our $Cd_3As_2$ SQUIDs. As shown in Fig. 2d, $\eta_c$ in SQUID1 displays an even function of the B fields around B = 0 for the positive and negative B fields. This is very different from the Meissner effect induced SDE, where the efficiency is an odd function of the B fields [61].

Time reversal symmetry requires $I_{+c}(+B) = |-I_{-c}(-B)|$. Therefore, as long as the SQUID does not break time reversal symmetry, no diode effect can be present even when inductance effects are taken into account, as discussed in [81,82]. In the SM [70], we provide a detailed theoretical calculation explicitly showing that even in the dynamical regime, when inductance effects are expected to be more relevant, the voltage-current (V-I) characteristic is symmetric when B=0, and its asymmetry is odd with respect to B when B is not zero, as required when the system does not spontaneously break time reversal symmetry.

To unravel the origin of the JDE observed in our SQUIDs, we notice that, as discussed in Refs. [20,21], when time-reversal is broken, an asymmetry between the transparency of the superconducting channels carrying the current across the SQUID is sufficient to explain the presence of a superconducting diode effect. The asymmetry between the JJ arms of our SQUIDs in quantities such as the sinusoidal form of the current phase relationship (CPR), self-inductances, critical currents $I_c$, and normal state resistances $R_n$, is sufficient to explain the presence of a JDE when B is not zero.

The zero-field JDE observed in SQUID2 requires further elucidation. In the absence of an external magnetic field, TRS must be broken in another way to realize a JDE. We propose that the Josephson coupling between a conventional superconductor (i.e., aluminum) and a two-band superconductor such as $Cd_3As_2$ can lead to TRS-breaking state characterized by a nonzero (mod $\pi$) phase difference between the two superconducting gaps as previously identified in Refs. [72, 83]. In our system superconducting correlations are induced in the bands (band 1 and 2) of $Cd_3As_2$ by the superconductivity in Al via the proximity effect. The coupling of band 1 and 2 to Al, a standard s-wave superconductor, favors order parameters in the two bands with no relative phase difference. Such "alignment" of the superconducting order parameters in the two bands can be frustrated if spin-orbit coupling favors pairing correlations in the two bands with a phase difference ~$\pi$. This situation is analogous to the one of two-band superconductors with a negative inter-band Josephson coupling [84]. In these conditions the phases of the superconducting order parameters in band 1 and 2 will have a "canted" equilibrium configuration with the phase of the order parameter in band 1 $\theta_1=\pm\theta_0$ and the phase of the order parameter in band 2 $\theta_2=-(\pm\theta_0)$, with $0< \theta_0< \pi$. The two possible states form a time-reversed pair, with each state in the pair breaking TRS. In a JJ based on Al/ $Cd_3As_2$, as the one shown in Fig.5(a), when the left side and right side of the junction are in different states of the time-reversed pair, see Fig. 5(a), the junction breaks the overall time reversal symmetry. In this situation the equilibrium phase difference, $\Delta\theta$, between the left and right lead of the JJ is non-zero for both bands: $\Delta\theta_1=\pm2\theta_0$, $\Delta\theta_2=-(\pm2\theta_0)$. The value of $\theta_0$, and therefore of $\Delta\theta_i$, depends on the coupling between Al and $Cd_3As_2$. Details of the interface between Al and $Cd_3As_2$, and the width and thickness of the Al layer, affect such coupling and therefore the value of $\theta_0$. For SQUIDs for which the values of $\theta_0$ in the two JJs are different a diode effect will be present even when no external magnetic field is present [85,86]. We expect that such configuration of phases might be realized in SQUID2 giving rise to the observed zero-field diode effect.

The fact that SQUID1 does not exhibit a diode effect in the absence of an external magnetic field can be due to the fact that for such SQUID, in the ground state, $\Delta\theta_1$ ($\Delta\theta_2$) is to good approximation the same in both JJs forming the SQUIDs. Another possibility is that for such SQUID the two superconducting channels have very different critical currents, for instance $I_{c1} >> I_{c2}$. The suppression of the diode effect for $I_{c1}>>I_{c2}$ can be illustrated with a minimal model for an asymmetric SQUID diode [21]: $I_1 = I_{c1}\sin\varphi$; $I_2 = I_{c2}\sin\varphi + I'_{c2}\sin 2\varphi$. The CPR of the SQUID in the absence of inductance is [21]

$$J(\varphi) = \sqrt{I_{c1}^2 + I_{c2}^2 + 2I_{c1}I_{c2}\cos\hat{\Phi}}\,\sin\varphi + I'_{c2}\sin\left(2\varphi - \tilde{\Phi}\right),$$

where $\hat{\Phi} = 2\pi\Phi/\Phi_0$, $\tilde{\Phi} = \hat{\Phi} + 2\gamma$ and $\tan\gamma = \frac{I_{c1}-I_{c2}}{I_{c1}+I_{c2}}\tan\frac{\hat{\Phi}}{2}$. In the limit $I_{c1}>>I_{c2}$, $\tilde{\Phi}\approx 0$ which suppresses the JDE. Fig. 5b shows the dependence of JDE efficiency, $\eta$, on the ratio $I_{c2}/I_{c1}$ with a magnetic flux $\Phi_0/4$ threading the SQUID ring. It is clearly seen that $\eta$ is almost zero if $I_{c2}/I_{c1}$ ~ 0.1 and reaches a maximal value for $I_{c2}$ ~ $I_{c1}$.

In conclusion, our combined experimental and theoretical analysis demonstrate that the coupling of the superconducting phases between the surface and bulk states can break TRS and induce a zero-magnetic-field JDE in topological SQUIDs made of Dirac semimetal $Cd_3As_2$. Importantly, by utilizing this unique property and the geometry of SQUIDs, one can control the efficiency of JDE. This should provide a practical approach in superconducting digital electronics applications.

We thank A. Sharpe for his critical reading of the manuscript and suggestions and are grateful to I. Žutić for his helpful comments. The work at Sandia was supported by the LDRD program. J.J.C., W.P., and E.R. acknowledge support from DOE, Grant No. DE-SC0022245. A portion of this work was performed at the National High Magnetic Field Laboratory, which is supported by National Science Foundation Cooperative Agreement No. DMR-2128556 and the State of Florida. Device fabrication was performed at the Center for Integrated Nanotechnologies, a U.S. DOE, Office of BES, user facility. Sandia National Laboratories is a multimission laboratory managed and operated by National Technology and Engineering Solutions of Sandia LLC, a wholly owned subsidiary of Honeywell International Inc. for the U.S. DOE's National Nuclear Security Administration under Contract No. DE-NA0003525. This paper describes objective technical results and analysis. Any subjective views or opinions that might be expressed in the paper do not necessarily represent the views of the U.S. DOE or the United States Government.

Figure1:

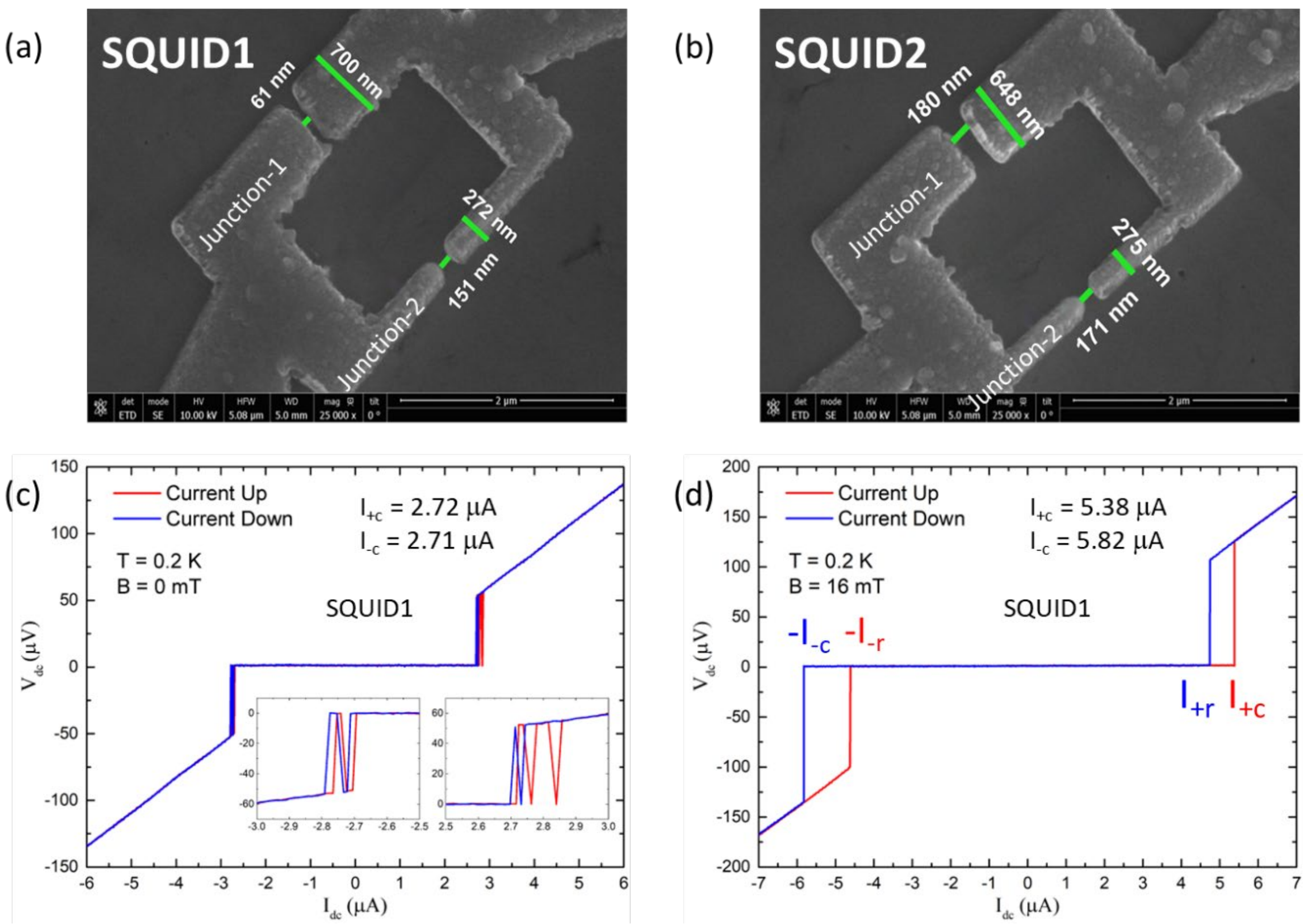

**Fig. 1: Asymmetric superconducting quantum interference devices and the I-V characterizations**. (a) and (b) SEM images of the two asymmetric SQUIDs studied. The black background represents $Cd_3As_2$ thin flakes. (c) Current-voltage (I-V) characteristics in SQUID1 at zero magnetic field B = 0. Multiple switching behaviors are observed in the critical and retrapping currents, see the bottom right inset. (d) I-V characteristics in SQUID1 at B = 16 mT. The position of $I_{+c}$, $-I_{-c}$, $I_{+r}$ and $-I_{-r}$ are marked. Herein, all four currents $I_{+c}$, $I_{-c}$, $I_{+r}$, and $I_{-r}$ are taken positive.

Figure 2:

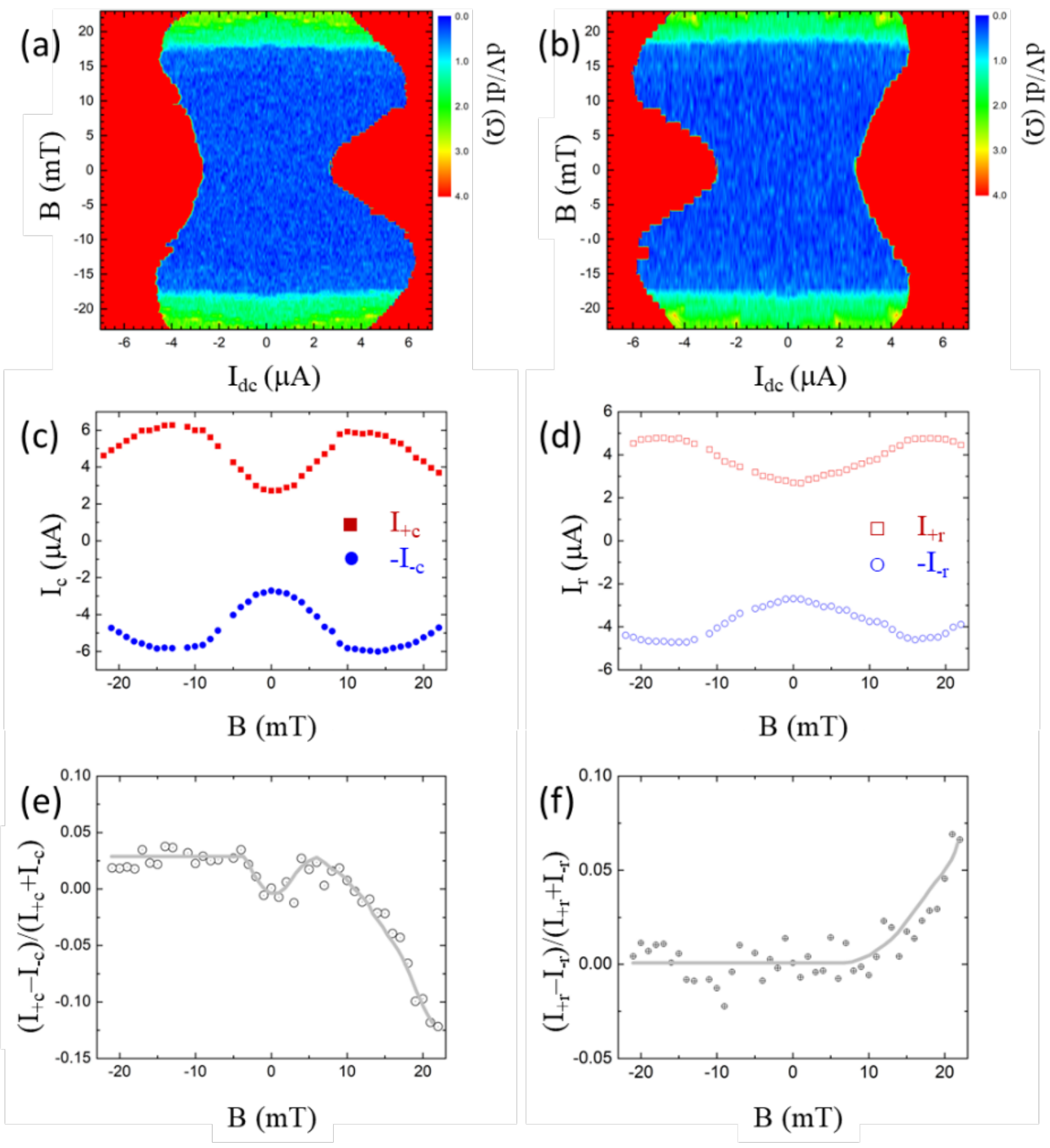

**Fig. 2. Magnetic field induced JDE in SQUID1**. (a) and (b) Differential resistance dV/dI as a function of DC current and B fields for two current sweeping directions. $I_{+c}$ and $-I_{-c}$, and the efficiency $\eta_c = (I_{+c}-I_{-c})/(I_{+c}+I_{-c})$ as a function of B field are shown in (c) and (e), respectively. $I_{+r}$ and $-I_{-r}$ (d) and $\eta_r = (I_{+r}-I_{-r})/(I_{+r}+I_{-r})$ (f) as a function of B field. The gray lines are a guide to the eye.

Figure 3:

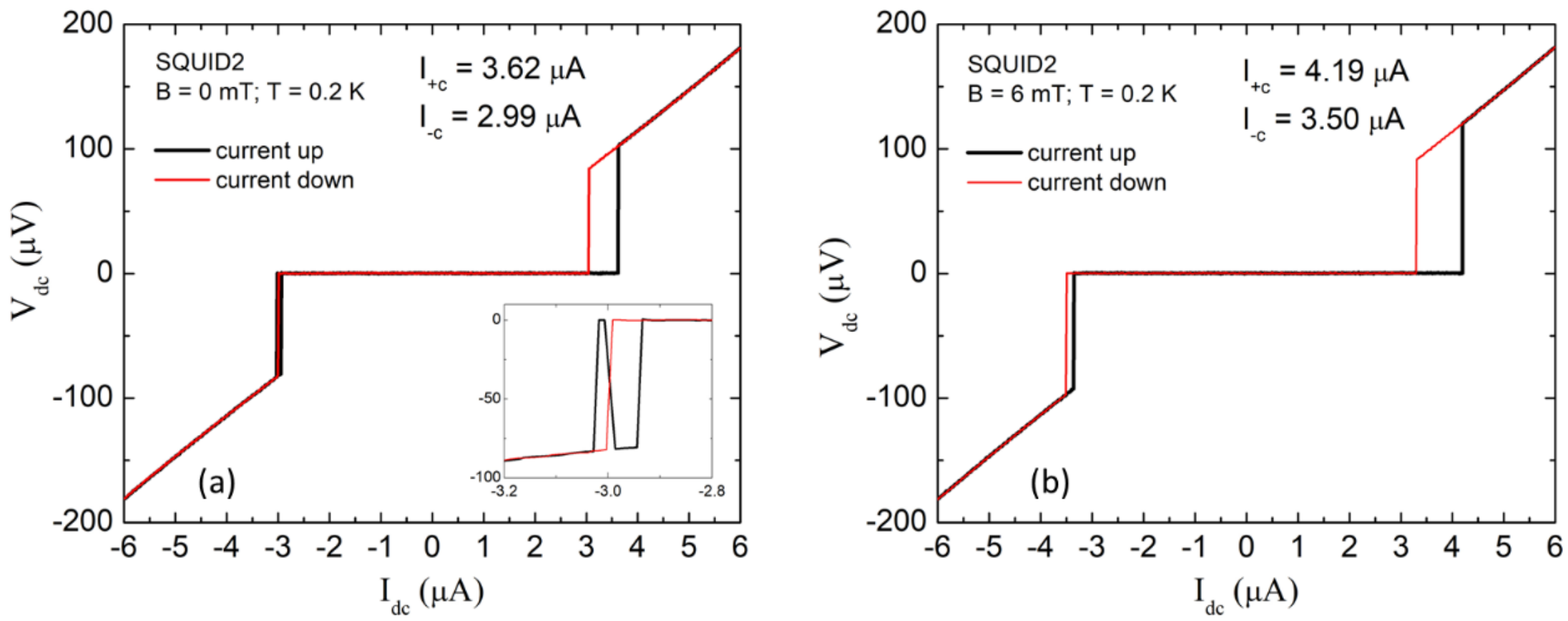

**Fig. 3 Zero-magnetic field JDE in SQUID2.** (a) I-V characteristics in SQUID2 at zero magnetic field. A large JDE is seen. $I_{+c}$ = 3.62 μA and $I_{-c}$ = 2.99 μA. This large JDE remains strong at a finite magnetic field of B = 6 mT (b).

Figure 4:

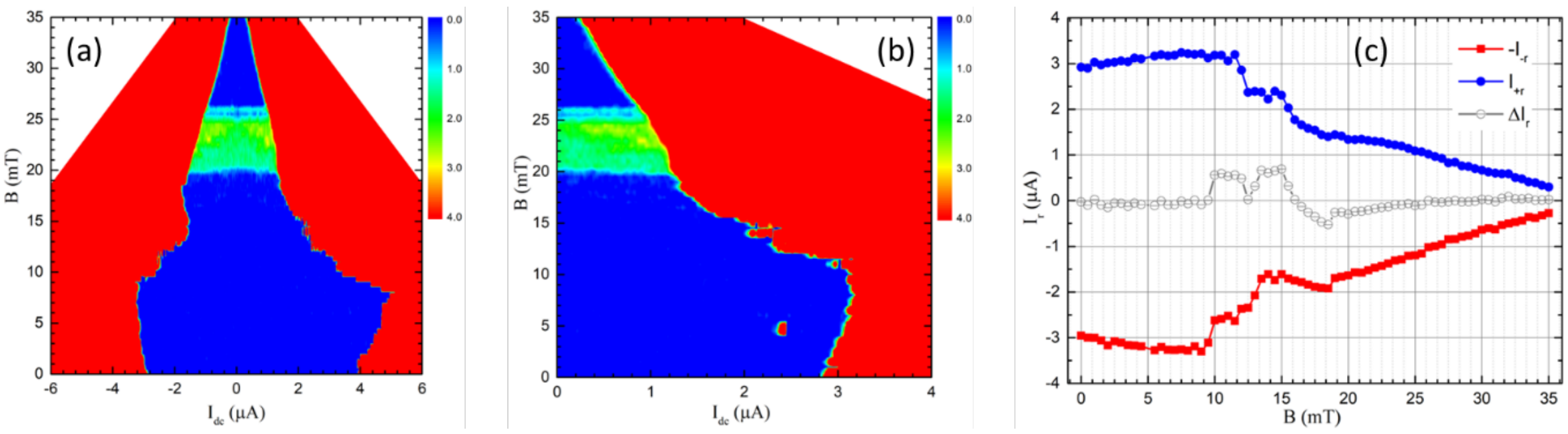

**Fig.4. JDE of retrapping currents in SQUID2.** (a) and (b) dV/dI as a function of DC current and magnetic field for two current sweeping directions. (c) $I_{+r}$, $-I_{-r}$, and $\Delta I_r$ as a function of B field.

Figure 5:

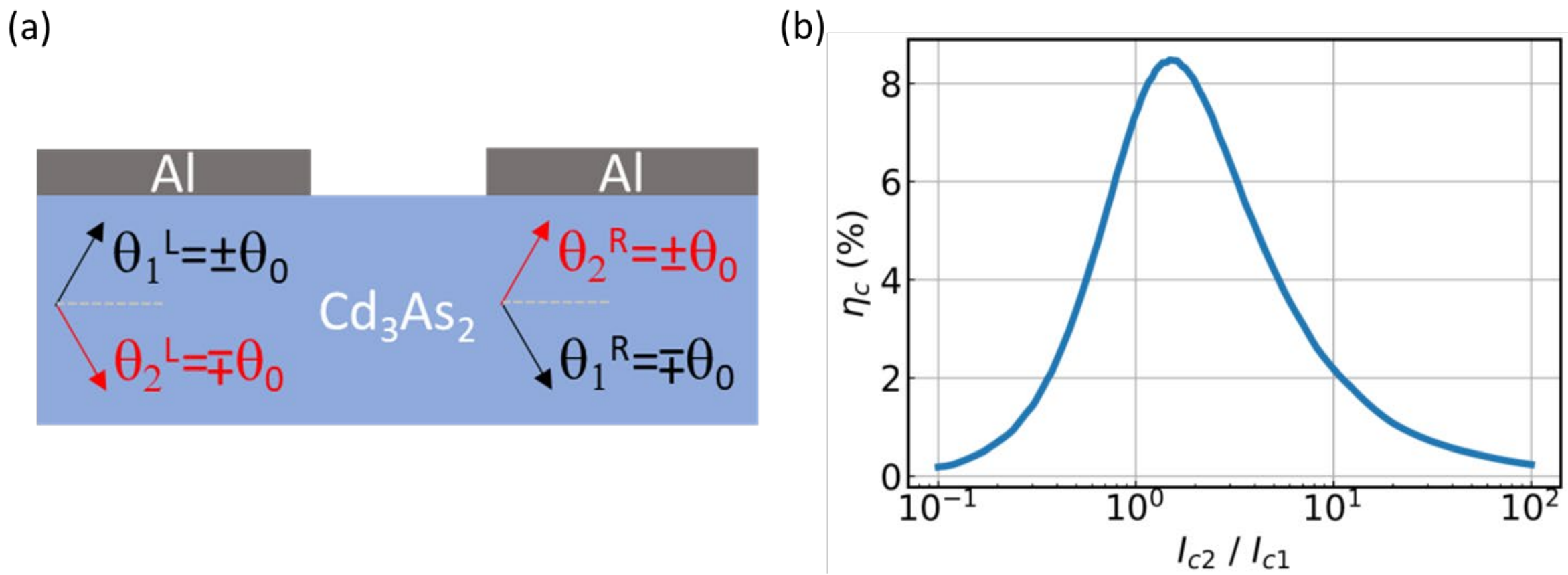

**Fig. 5 Theoretical analysis.** (a) Schematic representation of equilibrium phases of order parameters in a two-band superconductor Josephson junction. (b) Simulations of the efficiency ($\eta_c$) of JDE vs. the ratio of the critical currents in the two Josephson junction arms in an asymmetric SQUID.

## Supplementary Materials

# Time reversal symmetry breaking and zero magnetic field Josephson diode effect in Dirac semimetal $Cd_3As_2$-mediated asymmetric SQUIDs

W. Yu[1], J.J. Cuozzo[2], K. Sapkota[1], E. Rossi[3], D.X. Rademacher[1], T.M. Nenoff[1], W. Pan[2]

[1] Sandia National Labs, Albuquerque, NM 87185, USA

[2] Sandia National Labs, Livermore, CA 94551, USA

[3] Department of Physics, College of William and Mary, VA 23187, USA

### Device fabrication

A mechanical exfoliation method is used to obtain thin $Cd_3As_2$ flakes [72,73] of ~ 200 nm thickness from an initial bulk ingot material, synthesized via a chemical vapor deposition method [87]. These $Cd_3As_2$ thin flake are then deposited on a Si/$SiO_2$ substrate with a 1 μm thick $SiO_2$ layer. Next, e-beam lithography is used to define the SQUID structure. Aluminum (Al) superconducting electrodes of 300 nm thick are deposited using the e-beam evaporation method. To enhance adhesion, a thin layer of 5nm Ti is deposited before the Al layer.

In our device fabrication, we do not perform any specific surface treatments, such as argon milling, prior to Al deposition. In order to minimize the oxidization issue, our experimental device fabrication procedures are designed as follows. Immediately after mechanical exfoliation, we cover the surface of a $Cd_3As_2$ thin flake with PMMA. After e-beam writing and PMMA develop, we immediately move the device to an e-beam evaporator and pump overnight prior to Al deposition. Past published accountings of this procedure indicate that this device fabrication recipe is successful. For example, in our 2018 published paper (Ref. [72]), we show high-resolution TEM images of a high-quality interface in the Josephson junction using this device fabrication recipe.

Two SQUIDs are fabricated. Their SEM images are shown in Fig. 1. Overall, the two SQUIDs have similar dimensions (~ 2 μm × 1.3 μm) except that in SQUID1 the respective lengths of the two JJ arms differ considerably, while in SQUID2 they are roughly equal. In SQUID1, the size of the narrow junction (Junction-1) is ~ 151nm in length and ~ 272 nm in width, while for the wide junction (Junction-2) it is ~ 61 nm and ~ 700 nm, respectively. In SQUID2, the sizes for Junction-1 and Junction-2 are ~ 171 nm × 275 nm and ~ 180 nm × 648 nm, respectively.

Based on the geometry of our devices, a period of 0.8 mT is expected. This value, though slightly smaller than the step size of 1 mT in the measurements, is larger than the resolution (0.1 mT) of the superconducting magnet used.

Cautions are taken during the device fabrication to ensure a good interface between Al and $Cd_3As_2$, as demonstrated in our previous high resolution TEM studies [72]. This high-quality interface is also manifested by the proximity [71] induced supercurrent states in $Cd_3As_2$ JJs [72-76].

We note here that in Hall-bar like devices fabricated by using the same single crystal as the one used in the present work, the carrier density was determined to be $n_{2D} \sim 1.5 \times 10^{13}$ cm$^{-2}$ and $n_{3D} \sim 7.5 \times 10^{17}$ cm$^{-3}$ in the previous study [88]. Since the starting material is the same in both studies, we believe that the carrier density for our SQUIDs is in the same order.

**Differential resistance Measurements**:

To measure the differential resistance, a small AC current (dI = 2 nA) is added to the DC current between the two superconducting electrodes. A low frequency (~ 11Hz) phase lock-in amplifier technique is then used to measure the voltage drop (dV) between the same superconducting electrodes, from which the differential resistance dV/dI is obtained.

For the dV/dI measurements, first, dV/dI is measured with the DC current sweeping from +7 μA to -7 μA at various B fields. The same measurements are then repeated at the same B fields, but now the DC current sweeping from – 7 μA to + 7 μA.

**Temperature depedence of the junction resistance in SQUID1**

Fig. S1 shows the junction resistance of SQUID1 as a function of temperature (T) at zero magnetic field. The resistant is about 25 Ω at high T. The sharp drop at $T_c$ ~ 1.2K is due to the onset of a superconducting transition in Al electrodes. The resistance continues to decrease and reaches the zero-resistance state below T ~ 0.9K.

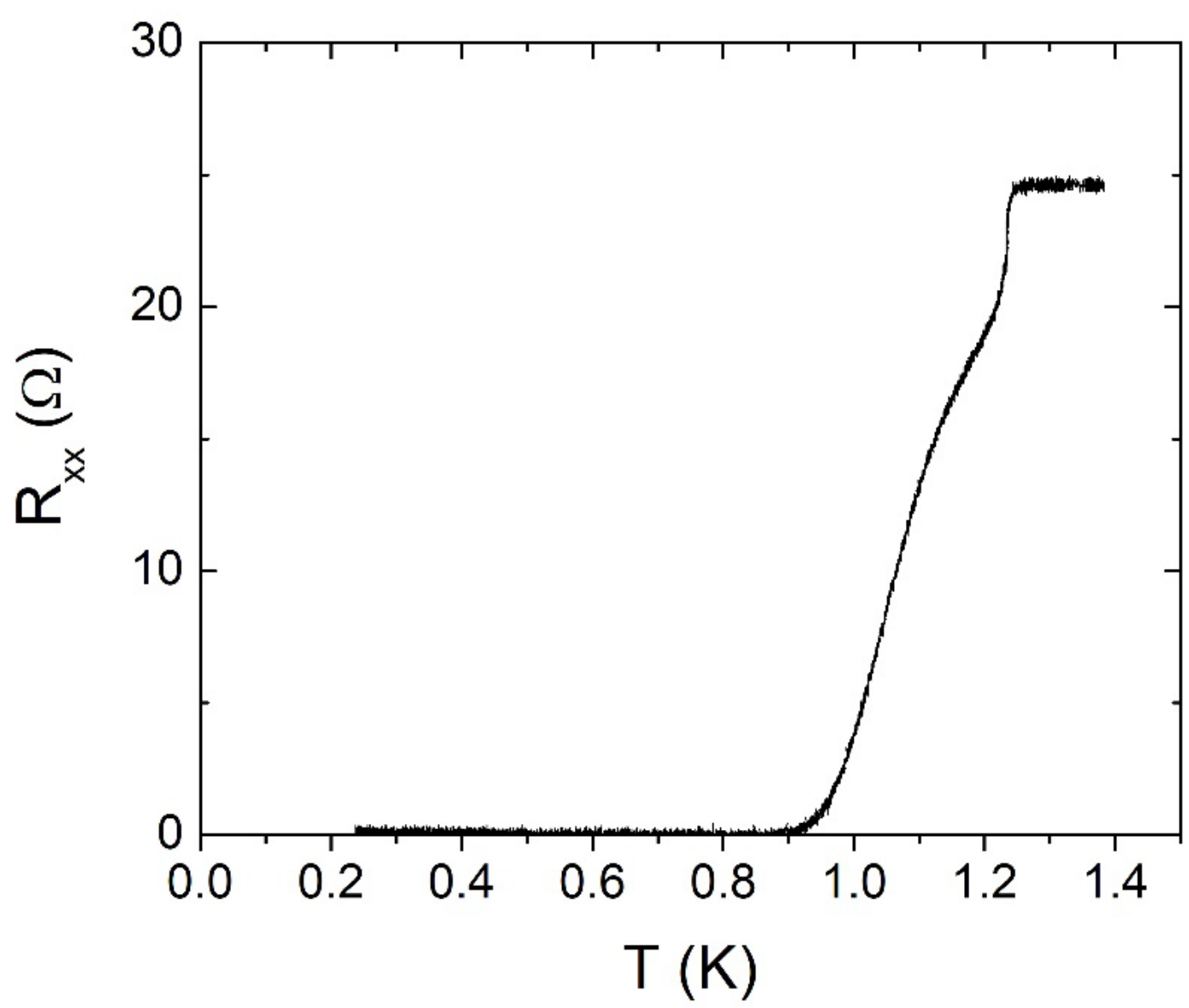

**Fig. S1.** Temperature dependence of the junction resistance in SQUID1.

**Evaluation of interface quality between Al and $Cd_3As_2$ in SQUID1 and SQUID2**

From the slope of the I-V curve in the high current regime, a normal state resistance of $R_n$ ~ 25 Ω is extracted. The $eI_cR_n$ product then gives a characteristic energy scale of ~ 68 μeV. This value is about half of the Al superconducting transition temperature of 1.2K (or an energy gap of Δ = 135 μeV, assuming $\Delta = 1.76K_BT_c$), indicating that a good interface has been achieved in SQUID1.

In SQUID2, the normal resistance is $R_n \sim 34\ \Omega$, and the average critical current is $(I_{+c}+I_{-c})/2 \sim$ 3.1 μA. So, the characteristic energy scale $eI_cR_n \sim$ 112 μeV. This is similar to the Al superconducting transition temperature of 1.2K (or an energy ga of $\Delta$ = 135 μeV), indicating that a better interface has been achieved in SQUID2. We caution here that we did not measure the T-dependence of junction resistance in SQUID2. As a result, the exact Al superconducting transition temperature is not known. However, we believe that $T_c$ in SQUID2 should be very close to that in SQUID1, as the two SQUIDs are fabricated at the same time using the same parameters.

**Theoretical calculation of the circuit inductance induced SDE/JDE**

Consider a SQUID described by the phenomenological resistively shunted junction model with the fluxoid quantization condition and conservation of charge:

$$\begin{cases} I_B = I_1 + I_2 \\ \phi_2 - \phi_1 = \dfrac{2\pi}{\Phi_0}\,\Phi_{tot} = \dfrac{2\pi}{\Phi_0}\,(L_1 I_1 - L_2 I_2 + \Phi_e) \\ I_k = \dfrac{\hbar}{2eR_k}\dfrac{d\phi_k}{dt} + I_{s,k}(\phi_k), \qquad k = 1,2. \end{cases}$$

Here $I_B$ is the current bias across the SQUID, $I_{1,2}$ are the currents in each SQUID arm, $\phi_{1,2}$ are the phases across each SQUID arm, $L_{1,2}$ are the inductances of each SQUID arm, $\Phi_e$ is the external magnetic flux threading the SQUID ring, $R_{1,2}$ are the normal resistances of each SQUID arm, and $I_{s,1,2}$ are the supercurrent channels of each SQUID arm. If we assume that the intrinsic gaps of each of the junctions in the SQUID are same ($I_{c1}R_1 = I_{c2}R_2$), then we can solve for a system of two equations for two phase variables $\phi_A = (\phi_1 + \phi_2)/2$ and $\psi = (\phi_2 - \phi_1)/2$ using unitless variables

$$\begin{cases} \frac{\hat{I}_B}{2} = \alpha \frac{d\phi_A}{d\tau} + \frac{I_{s,1}(\phi_1) + I_{s,2}(\phi_2)}{2I_{c1}} + \epsilon \frac{d\psi}{d\tau} \\ \beta_L \left( \widehat{\Phi}_e + \widehat{\Phi}_B - \frac{\psi}{\pi} \right) = \alpha \frac{d\psi}{d\tau} + \frac{I_{s,2}(\phi_2) - I_{s,1}(\phi_1)}{2I_{c1}} + \epsilon \frac{d\phi_A}{d\tau}. \end{cases}$$

Here $\widehat{\Phi}_B \propto (L_1 - L_2) I_B$ is a magnetic flux associated with an asymmetry in inductances, $\alpha, \beta_L, \epsilon$ are unitless constants, $\tau$ is unitless time, and hats denote unitless variables. Now, the reciprocity of the SQUID can be characterized by the symmetry of the time-averaged voltage across the device $\langle V \rangle_\tau \propto \left\langle \frac{d\phi_A}{d\tau} \right\rangle_\tau$ . Writing the phase difference in terms of its equilibrium and dynamic parts, $\psi(\tau) = \tilde{\psi}(\tau) + \psi_0$ where $\psi_0 = const.$, we can solve for $\langle V \rangle_\tau$ up to a constant factor:

$$\langle V \rangle_\tau \propto \left\langle \frac{\hat{I}_B}{2} - \frac{I_{s1}(\phi_1) + I_{s2}(\phi_2)}{2I_{c1}} - \frac{\epsilon \beta_L}{\alpha} \left( \widehat{\Phi}_e + \widehat{\Phi}_B - \frac{\psi}{\pi} \right) - \frac{\epsilon}{2I_{c1}\alpha} \left( I_{s1}(\phi_1) - I_{s2}(\phi_2) \right) \right\rangle_\tau$$

To observe a diode effect, the VI characteristic must be asymmetric i.e. $\langle V \rangle_\tau^{(+)} \neq -\langle V \rangle_\tau^{(-)}$ where $\langle V \rangle_\tau^{(\pm)} = \langle V(\pm\phi_1, \pm\phi_2, \pm I_B) \rangle_\tau$. In a standard SQUID where $I_s(\phi) = I_c \sin\phi$ and $\psi_0 = 0$, the VI characteristic can be asymmetric only if there exists a finite magnetic flux $\widehat{\Phi}_e$ threading the SQUID. Furthermore, since $\widehat{\Phi}_e$ is the only asymmetric term (linear in magnetic field), this implies the non-reciprocity is an odd function of the magnetic field.

**Re-entrant supercurrent states at high magnetic fields in SQUID2**

As shown in Figs. 4a and 4b, the supercurrent states disappear around 19 mT, but they reappear when B reaches ~ 27 mT. Fig. S2 shows three traces at B = 19, 24, and 29 mT. Around the zero bias, dV/dI = 0 at 19 and 29 mT but becomes finite at B = 24 mT. The exact origin of this re-entrant superconducting states is under further investigation. However, we notice that re-entrant superconducting states were previously observed in other materials systems, there it was believed to be either due to phase-slip centers or vortex motion-related [89,90].

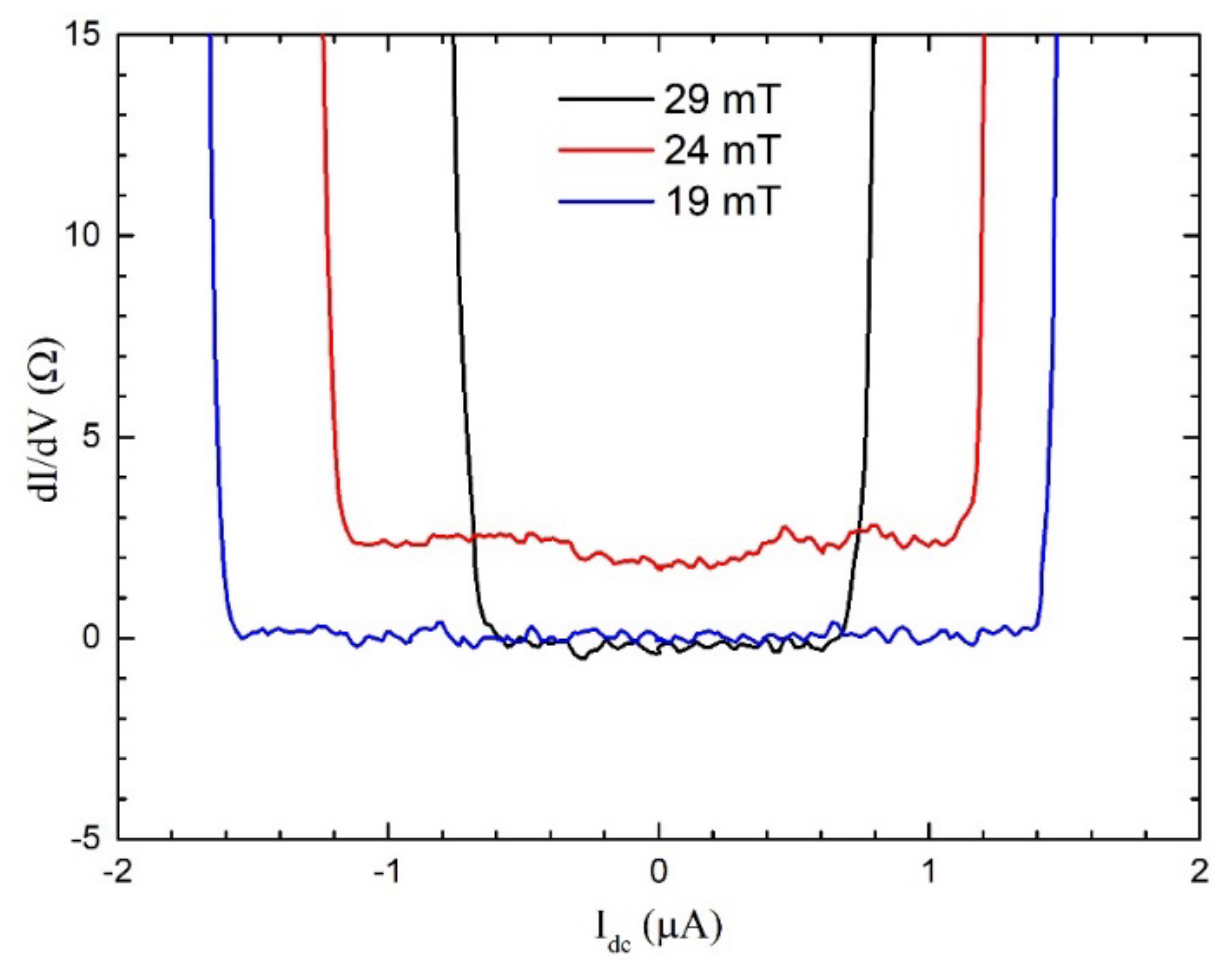

**Fig. S2.** dV/dI as a function of DC current at three selected magnetic fields B = 19, 24, and 29 mT. dV/dI = 0 around the zero bias at B = 19 and 29mT but assumes a finite value at B = 24 mT.